\documentclass[aps,pre,twocolumn,showpacs,longbibliography,10pt,nobalancelastpage]{revtex4-1}

\usepackage{amsmath,amssymb}
\usepackage{graphicx}
\usepackage{xcolor}
\usepackage{mathrsfs}
\usepackage{newtxtext}
\usepackage[slantedGreek,cmintegrals]{newtxmath}
\usepackage{stmaryrd}
\usepackage{hyperref}
\usepackage{enumitem}
\usepackage{accents}
\hypersetup{colorlinks,linkcolor={blue!60!black},citecolor={blue!60!black},urlcolor={blue!60!black}}

\renewcommand{\vec}[1]{\boldsymbol{#1}}

\renewcommand{\pi}{\uppi}

\DeclareMathAlphabet{\mathcal}{OMS}{cmsy}{m}{n}
\DeclareMathAlphabet{\mathcalbf}{OMS}{cmsy}{b}{n}
\DeclareMathAlphabet{\mathbfsfit}{\encodingdefault}{\sfdefault}{b}{it}
\DeclareMathAlphabet{\mathbfsf}{\encodingdefault}{\sfdefault}{b}{n}

\renewcommand{\leq}{\leqslant}
\renewcommand{\geq}{\geqslant}

\begin{document}

\title{Nonlinear and Nonlocal Elasticity in Coarse-Grained Differential-Tension Models of Epithelia}
\author{Pierre A. Haas}
\email{P.A.Haas@damtp.cam.ac.uk}
\author{Raymond E. Goldstein}
\email{R.E.Goldstein@damtp.cam.ac.uk}
\affiliation{Department of Applied Mathematics and Theoretical Physics, Centre for Mathematical Sciences, \\ University of Cambridge, 
Wilberforce Road, Cambridge CB3 0WA, United Kingdom}
\date{\today}%
\begin{abstract}
The shapes of epithelial tissues result from a complex interplay of contractile forces in the cytoskeleta of the cells in the tissue, and adhesion forces between them. A host of discrete, cell-based models describe these forces by assigning different surface tensions to the apical, basal, and lateral sides of the cells. These differential-tension models have been used to describe the deformations of epithelia in different living systems, but the underlying continuum mechanics at the scale of the epithelium are still unclear. Here, we derive a continuum theory for a simple differential-tension model of a two-dimensional epithelium and study the buckling of this epithelium under imposed compression. The analysis reveals how the cell-level properties encoded in the differential-tension model lead to linear, nonlinear as well as nonlocal elastic behavior at the continuum level.
\end{abstract}

\maketitle

\section{Introduction}\enlargethispage{5mm}
Intercellular adhesion proteins and cortical actin networks 
are well established as regulators of cell surface mechanics, and hence of the deformations of epithelia during morphogenesis~\cite{lecuit07}. 
Ever since the seminal work of Odell {\it et al.} \cite{OdellOster}, these cellular components have therefore underlain mathematical models of epithelia \cite{Leptin,Baker}. One large class of such models are differential-tension models \cite{derganc09,hocevar12,krajnc13,hannezo14,rauzi15,krajnc15,storgel16,krajnc18}, in which cell polarity, cell-cell adhesion properties and the actomyosin network induce different surface tensions in different sides of the discrete cells. Coupling the mechanical models describing epithelial deformations to models of the intracellular biochemistry is a key challenge in the field~\cite{howard11}, but some progress has recently been made by coupling the differential-tension model of Ref.~\cite{hannezo14} to the diffusion of a `mechanogen' that induces contractility~\cite{dasbiswas18}.

While the differential-tension models can quantitatively reproduce the morphology of epithelial folds in many different living systems~\cite{storgel16}, it is likely that, in general multilayered epithelia, the formation of epithelial folds must be ascribed to a combination of these intra-epithelial stresses and differential growth of different parts of the tissue. Models based on the latter only have for example been invoked to describe, at the scale of the epithelium, the formation of cortical convolutions in the brain~\cite{richman75,tallinen13,manyuhina14,budday15,tallinen16,lejeune16} and of the intestinal villi~\cite{hannezo11,savin11,shyer13}, the `fication' of which lends itself to the pun that gave Ref.~\cite{shyer13} its title. The coarse-grained limit of the differential-tension models at this scale is less well-studied, however, and this continuum limit is the topic of this paper.

\begin{figure}[b]
\centering\includegraphics{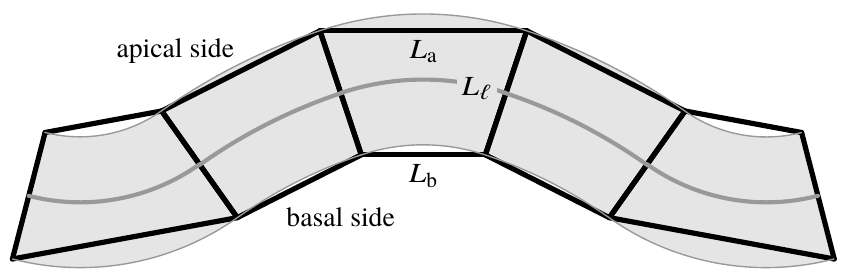}
\caption{Model epithelium. Isosceles trapezoidal cells of apical and basal bases $L_{\mathrm{a}}$ and $L_{\mathrm{b}}$ are connected along their lateral sides, which have length $L_\ell$. The continuous line and shaded area provide a cartoon of the continuum limit.}
\label{fig:model}
\end{figure}

We shall focus on the most basic setup of these differential-tension models~\cite{derganc09,krajnc13,krajnc15} in an epithelial monolayer: the apical, basal, and lateral sides of the cells have respective areas $A_{\mathrm{a}}$, $A_{\mathrm{b}}$, and $A_\ell$. The internal state of the cells induces different surface tensions $\Gamma_{\mathrm{a}}$, $\Gamma_{\mathrm{b}}$, and $\Gamma_\ell$ in the apical, basal, and lateral sides of the cells, respectively. The energy of a single cell therefore reads
\begin{align}
E=\Gamma_{\mathrm{a}}A_{\mathrm{a}}+\Gamma_{\mathrm{b}}A_{\mathrm{b}}+\dfrac{\Gamma_{\ell}}{2}A_{\ell},\label{eq:singlecelleE0}
\end{align}
where the factor of $1/2$ has been introduced for mere convenience~\cite{krajnc13,krajnc15}. The theory can be extended to incorporate additional physics such as a basement membrane~\cite{storgel16} or a confining vitelline membrane~\cite{hocevar12}. In a full three-dimensional setup, this leads to the study of the shapes of prism-shaped cells~\cite{hannezo14,krajnc18}. Here, we shall restrict to the two-dimensional setup~\cite{krajnc13,krajnc15} of an epithelium consisting of isosceles trapezoidal cells of parallel apical and basal sides of lengths $L_{\mathrm{a}}$ and $L_{\mathrm{b}}$. These trapezoids are joined up across their lateral sides, which have equal length $L_{\mathrm{\ell}}$ (Fig.~\ref{fig:model}). Since there are two such lateral sides, the cell energy (\ref{eq:singlecelleE0}) reduces to
\begin{align}
E= \Gamma_{\mathrm{a}}L_{\mathrm{a}}+\Gamma_{\mathrm{b}}L_{\mathrm{b}}+\Gamma_{\ell}L_{\ell},\label{eq:singlecellE}
\end{align}
per unit extent in the third dimension~\cite{krajnc13,krajnc15}. A different two-dimensional limit is obtained by averaging over the thickness of the cell sheet and describing in-plane deformations only. Such models, termed area- and perimeter-elasticity models, have been studied extensively~\cite{hufnagel07,farhadifar07,hilgenfeldt08,bi15,bi16}.

The simplest problem in the mechanics of elastic rods is their Euler buckling under applied forces~\cite{landaulifshitz}; it is therefore meet to ask how the buckling behavior of an active material such as this model epithelium differs from that of an elastic rod. This problem was considered in Ref.~\cite{krajnc15}, where the continuum limit of Eq.~(\ref{eq:singlecellE}) was mapped to Euler's Elastica equation~\cite{* [] [{ App. A, pp. 546--557; Chap. 6.4, pp. 184--192 and App. D, pp. 571--581.}] audolypomeau}. These calculations, complementing simulations of the discrete model in Ref.~\cite{krajnc13}, were phrased in terms of spontaneous buckling of the epithelium, but an additional compressive or extensile force is required to produce these deformations, making the analysis of Ref.~\cite{krajnc15} more appropriate to the present setup of buckling under imposed forces. Moreover, the analysis of Ref.~\cite{krajnc15} is not completely consistent with the discrete model, since it does not impose the condition that the trapezoidal cells match up exactly along their lateral sides.

\begin{figure*}
\centering\includegraphics{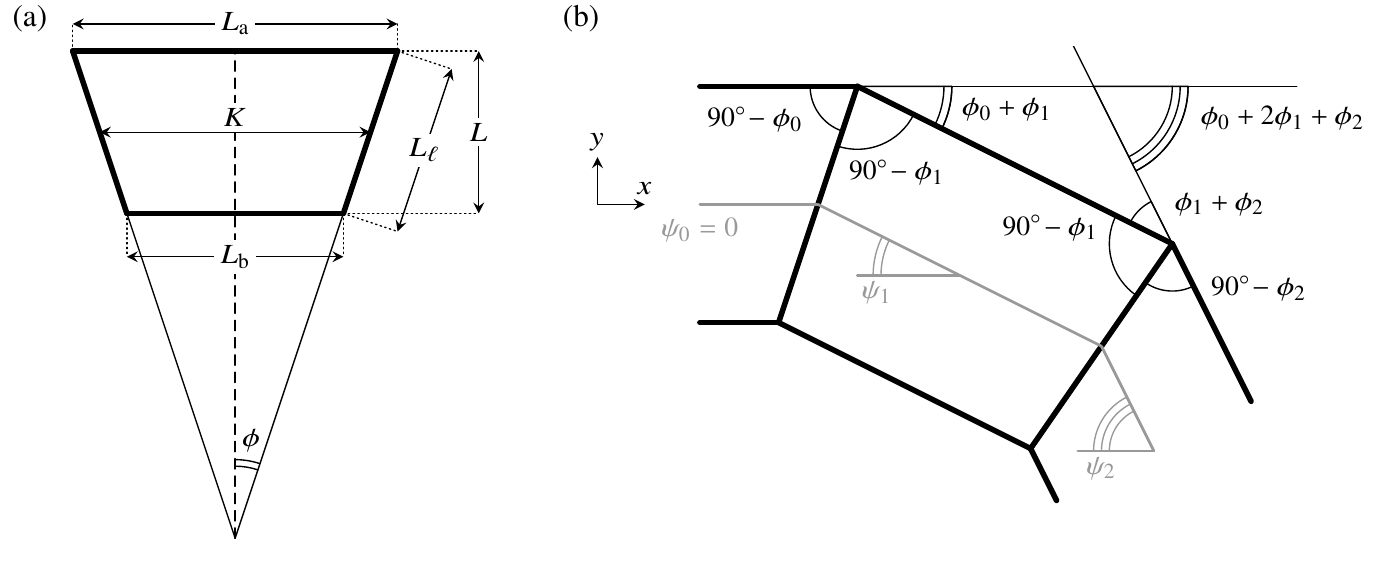}
\caption{Cell Geometry. (a)~Geometry of a single isosceles trapezoidal cell of mean base~$K$ and height~$L$, and sidelengths $L_{\mathrm{a}},L_{\mathrm{b}},L_\ell$, the lateral sides being at an angle $2\phi$ to each other. (b)~Definition of the tangent angle $\psi$ of the midline below the horizontal. The geometry of contiguous cells defines the relation between $\phi$ and $\psi$, as expressed in Eq.~(\ref{eq:psigeo}).}
\label{fig:geo}
\end{figure*}

In this paper, we perform a consistent asymptotic expansion of the discrete geometry of this model, revealing nonlinear and nonlocal elastic terms in the continuum limit. We then analyze the buckling behavior of the continuum model under imposed compression analytically and numerically. 

\section{Continuum Model}
\subsection{Single-cell Energy}
To obtain the energy of a single cell, we express the sidelengths of the trapezoidal cells in terms of their mean base~$K$, their height~$L$, and the angle $2\phi$ that their lateral sides make with each other (Fig.~\ref{fig:geo}a):
\begin{subequations}\label{eq:cellgeom}
\begin{align}
L_{\mathrm{a}}&=K+L_\ell\sin{\phi},\label{eq:cellgeoma}\\
L_{\mathrm{b}}&=K-L_\ell\sin{\phi},\label{eq:cellgeomb}\\
L&=L_\ell\cos{\phi}.\label{eq:cellgeomc}
\end{align}
Incompressibility implies the cell area conservation constraint
\begin{align}
A_{\mathrm{c}}=KL. \label{eq:areacons}
\end{align}
\end{subequations}
Upon eliminating $K$ and $L$ using these relations, the energy of a single cell is expressed, from Eq.~(\ref{eq:singlecellE}) and as a function of $L_\ell$ and $\phi$, as
\begin{align}
E=(\Gamma_{\mathrm{a}}+\Gamma_{\mathrm{b}})\dfrac{A_{\mathrm{c}}}{L_\ell}\sec{\phi}+(\Gamma_{\mathrm{a}}-\Gamma_{\mathrm{b}})L_\ell\sin{\phi}+\Gamma_\ell L_\ell. 
\end{align}

\subsubsection{Non-dimensionalization}
We non-dimensionalize this expression by scaling lengths with the square root of the cell area and thus define the non-dimensional length of the lateral sides of the trapezoidal cells, $\smash{\lambda=L_\ell/A_{\mathrm{c}}^{1/2}}$. We further set $\smash{\ell=L/A_{\mathrm{c}}^{1/2}}$ and $\smash{\kappa=K/A_{\mathrm{c}}^{1/2}}$. Finally, following Ref.~\cite{krajnc15}, we introduce the parameters
\begin{align}
&\alpha=\dfrac{\Gamma_{\mathrm{a}}}{\Gamma_{\ell}}, &&\beta=\dfrac{\Gamma_{\mathrm{b}}}{\Gamma_{\ell}},&&\ell_0=\sqrt{\alpha+\beta},&&\delta=\alpha-\beta.
\end{align}
We note that $\ell_0$ is the (uniform and non-dimensionalized) thickness of the epithelium in the flat configuration. Hence $s_0=1/\ell_0$ is the (non-dimensionalized) width of a single cell (i.e. its arclength in the flat configuration). The non-dimensional energy $\smash{e=E/\Gamma_\ell A_{\mathrm{c}}^{1/2}}$ of a single cell is therefore
\begin{align}
e=\dfrac{\ell_0^2}{\lambda}\sec{\phi}+\lambda\bigl(\delta\sin{\phi}+1\bigr).\label{eq:nondime}
\end{align}
Without loss of generality, we assume that $\delta>0$ in what follows, so, for a single cell, $\phi<0$ is energetically favorable.
\subsubsection{Transition to Constricted Cells}
The transition to constricted triangular cells is a geometric singularity in the discrete model. These triangular cells arize as limiting cases of the trapezoidal cells when $L_{\text{a}}=0$ or $L_{\text{b}}=0$. Using Eq.~(\ref{eq:cellgeom}), the conditions $L_{\text{a}},L_{\text{b}}\geqslant 0$ reduce to 
\begin{align}
\lambda^2\leqslant\dfrac{1}{\cos{\phi}\left|\sin{\phi}\right|}=\dfrac{2}{\left|\sin{2\phi}\right|}. 
\end{align}

\subsection{Energy of an Epithelium}
In the continuum limit, we take $\phi$ to be a function of the arclength $s$ of the midline of the undeformed, flat epithelium. Summing over all the cells, we obtain the non-dimensional energy $\mathcal{E}$ of the epithelium,
\begin{align}
\mathcal{E}=\int{e(\phi)\,\ell_0\,\mathrm{d}s}=\ell_0^2\int{\left(\dfrac{\ell_0}{\lambda}\sec{\phi}+\dfrac{\lambda}{\ell_0}\bigl(1+\delta\sin{\phi}\bigr)\right)\,\mathrm{d}s}, 
\end{align}
By imposing an energy density equal to $1/s_0=\ell_0$ times the (non-dimensional) energy of a single cell and integrating with respect to the reference arclength in this manner, we have imposed local cell area conservation \footnote{In this respect, the present analysis differs from that of Ref.~\cite{krajnc15}, where the energy was expressed as an integral with respect to the arclength~$S$ in the deformed configuration, related in fact to $s$ by $\ell\,\mathrm{d} S=\ell_0\,\mathrm{d} s$. In Ref.~\cite{krajnc15}, a weaker global area conservation constraint for the cell sheet had therefore to be imposed separately.}. 

%To complete the setup of the model, we introduce the angle $\psi$ of the deformed midline of the epithelium \emph{below} the horizontal (Fig.~\ref{fig:geo}b), and express the energy in terms of $\psi$.  This is most
%usefully done in a particular scaling limit, as explained next.

The boundary conditions %on the epithelium 
are most naturally expressed in terms of the angle $\psi$ of the deformed midline of the epithelium \emph{below} the horizontal (Fig.~\ref{fig:geo}b). 
We therefore express the energy in terms of 
$\psi$. This is usefully done in the scaling limit $\ell_0\gg 1$ of a columnar epithelium, as explained next.

\subsubsection{Asymptotic Expansion}
%We make two further scaling assumptions for the asymptotic analysis:
%The core of the analysis is an asymptotic expansion in the large parameter $\ell_0\gg1$, corresponding to a columnar epithelium. We begin by making two scaling assumptions,
We make two further scaling assumptions:
\begin{align}
&\lambda=\mathcal{O}(\ell_0),&&\phi=\mathcal{O}\bigl(\ell_0^{-2}\bigr).
\end{align}
These scalings correspond to the regime $\ell\sim\ell_0$ and $\phi\ll 1$, where the cells deform but slightly from their equilibrium configuration. In this limit, Eq.~(\ref{eq:cellgeomc}) implies that $\lambda\sim\ell\sim\ell_0$. Further, area conservation (\ref{eq:areacons}) requires that $\kappa\sim1/\ell_0$, and thus, from Eqs.~(\ref{eq:cellgeoma},\ref{eq:cellgeomb}), we must have $\phi\lesssim \kappa/\lambda\sim 1/\ell_0^{2}$. The second scaling thus corresponds to the largest deformations allowed. We therefore introduce the parameter
\begin{align}
%\Lambda=\dfrac{\lambda}{\ell_0}=\mathcal{O}(1).
\Lambda=\lambda/\ell_0=\mathcal{O}(1).
\end{align}
We are now set up to relate $\phi$ and $\psi$, for which purpose we use the geometric relation
\begin{align}
\psi(s+ks_0)-\psi(s)&=\phi(s)+2\phi(s+s_0)+\cdots\nonumber\\
&\hspace{5mm}+2\phi\bigl(s\!+\!(k\!-\!1)s_0\bigr)+\phi(s\!+\!ks_0),\label{eq:psigeo} 
\end{align}
valid for any positive integer $k$, as sketched in Fig.~\ref{fig:geo}b. In Appendix~\ref{appA}, we show that, with our scaling assumptions, the continuum limit of this relation is
\begin{align}
\psi'(s)=\sum_{m=0}^\infty{\dfrac{2\mathcal{B}_{2m}}{(2m)!}\dfrac{\phi^{(2m)}(s)}{\ell_0^{2m-1}}}, \label{eq:dpsi}
\end{align}
wherein $\mathcal{B}_0=1,\mathcal{B}_1=-\frac{1}{2},\dots$ are the Bernoulli numbers (of the first kind)~\cite{bernoulli}. The next step is to invert this series, to express $\phi$ in terms of the derivatives of $\psi$. While we are not aware of any explicit expression for the coefficients of the inverted series, it is straightforward to invert the series order-by-order by substituting back and forth, and thus obtain
\begin{align}
\phi(s)=\dfrac{\psi'(s)}{2\ell_0}-\dfrac{\psi'''(s)}{24\ell_0^3}+\dfrac{\psi^{(\text{v})}(s)}{240\ell_0^5}+\cdots,\label{eq:phiexp} 
\end{align}
where dashes denote differentiation with respect to $s$. In Ref.~\cite{krajnc15}, only the first term of this expansion was obtained. Inclusion of the second term will enable us to analyze the buckling behavior of the epithelium in what follows.

\subsubsection{Shape Equations for the Buckled Epithelium}
We describe the shape of the buckled epithelium by the coordinates $\bigl(x(s),y(s)\bigr)$ of the centreline of the epithelium, defined by the axes in Fig.~\ref{fig:geo}b. To derive the continuum equations describing the centreline, we begin by projecting the discrete geometry onto the axes,
\begin{subequations}
\begin{align}
%&x(s+s_0)-x(s)\nonumber\\
%&\hspace{5mm}= \dfrac{1}{2}\Bigl(\kappa(s)\cos{\psi(s)}+\kappa(s+s_0)\cos{\psi(s+s_0)}\Bigr),\\
x(s\!+\!s_0)-x(s)&= \dfrac{1}{2}\Bigl(\kappa(s)\cos{\psi(s)}+\kappa(s\!+\!s_0)\cos{\psi(s\!+\!s_0)}\Bigr),\\
y(s\!+\!s_0)-y(s)&= \dfrac{1}{2}\Bigl(\kappa(s)\sin{\psi(s)}+\kappa(s\!+\!s_0)\sin{\psi(s\!+\!s_0)}\Bigr).
\end{align}
\end{subequations}
Using $\kappa(s)=(\ell_0\Lambda)^{-1}\sec{\phi(s)}$ and expanding these equations order-by-order in inverse powers of $\ell_0$ using Eq.~(\ref{eq:phiexp}), we obtain, after a considerable amount of algebra~\footnote{These and other expansions in this paper were carried out with the help of \textsc{Mathematica} (Wolfram, Inc.) to facilitate the manipulation of complicated algebraic expressions.},
\begin{align}
&\Lambda\dfrac{\mathrm{d}x}{\mathrm{d}s}=f\cos{\psi}-g\sin{\psi},&&\Lambda\dfrac{\mathrm{d}y}{\mathrm{d}s}=f\sin{\psi}+g\cos{\psi},
\end{align}
where
\begin{subequations}\label{eq:fgdef}
\begin{align}
f&=1+\dfrac{{\psi'}^2}{24\ell_0^2}+\dfrac{7{\psi'}^4+144{\psi''}^2+32\psi'\psi'''}{5760\ell_0^4}+\mathcal{O}\bigl(\ell_0^{-6}\bigr),\label{eq:fdef}  \\
g&=\dfrac{\psi''}{12\ell_0^2}+\dfrac{87{\psi'}^2\psi''-2\psi^{\text{(iv)}}}{1440\ell_0^4} +\mathcal{O}\bigl(\ell_0^{-6}\bigr).
\end{align}
\end{subequations}
Integrating these differential equations yields the shape of the buckled epithelium. Deviations from the `standard' values $f=1$, $g=0$ arise at order $\smash{\mathcal{O}\bigl(\ell_0^{-2}\bigr)}$.
\subsubsection{Derivation of the Governing Equation}
We shall seek to describe buckled configurations of an epithelium of undeformed length $2\Sigma\gg s_0$. We change variables by introducing $\sigma=s/\Sigma$, use dots to denote differentiation with respect to $\sigma$, and define
\begin{align}
\Xi=\ell_0\Sigma\gtrsim\mathcal{O}(1).
\end{align}
Since $\Xi=\Sigma/s_0$, the number of cells in the epithelium is simply $N=2\Xi$. 

We shall seek buckled solutions with clamped boundary conditions and a prescribed relative compression $D$, so that
\begin{align}
&x(2)-x(0) = 2(1-D), && y(2) = y(0),
\end{align}
where the coordinates are now expressed relative to the scaled arclength $\sigma$. We shall restrict to symmetrically buckled configurations for which $\psi(\sigma)=-\psi(2-\sigma)$. The second condition above is then satisfied. We may further reduce the solution to the range $0\leq\sigma\leq 1$, with the condition of prescribed compression reading $x(1)-x(0)=1-D$. To minimize the energy of the epithelium at this imposed displacement, we therefore consider the Lagrangian
\begin{align}
\mathcal{L}&=\int_0^1{\left(\dfrac{\sec{\phi(\sigma)}}{\Lambda}+\Lambda\bigl(1+\delta\sin{\phi(\sigma)}\bigr)\right)\mathrm{d}\sigma}\nonumber\\
&\hspace{45mm}+\dfrac{\mu}{\Sigma}\int_0^1{\dot{x}(\sigma)\,\mathrm{d}\sigma}, 
\end{align}
where the Lagrange multiplier $\mu$ imposes the displacement condition and has the interpretation of a horizontal, compressive force. 

Upon substituting for $\phi$ using Eq.~(\ref{eq:phiexp}), expanding in inverse powers of $\Xi$, discarding terms that vanish upon integration, and integrating by parts, we find
\begin{widetext}
\begin{align}
\mathcal{L}=\int_0^1{\left[\Lambda+\dfrac{1}{\Lambda}+\dfrac{\dot{\psi}^2}{8\Lambda\Xi^2}-\dfrac{\delta\Lambda\dot{\psi}^3}{48\Xi^3}+\dfrac{5\dot{\psi}^4}{384\Lambda\Xi^4}+\dfrac{\ddot{\psi}^2}{48\Lambda\Xi^4}+\dfrac{\mu\cos{\psi}}{\Lambda}\left(1+\dfrac{\dot{\psi}^2}{8\Xi^2}+\dfrac{41\dot{\psi}^4}{1920\Xi^4}+\dfrac{\ddot{\psi}^2}{40\Xi^4}+\dfrac{\dot{\psi}\dddot{\psi}}{240\Xi^4}\right)+\mathcal{O}\bigl(\Xi^{-5}\bigr)\right]\mathrm{d}\sigma}.\label{eq:Lagrangian}
\end{align}
To analyze the dependence of the energy on the differential tension $\delta$, we must go beyond lowest order~\footnote{A non-trivial term arises already at order $\mathcal{O}\bigl(\Xi^{-2}\bigr)$ in the second part of Eq.~(\ref{eq:Lagrangian}) that imposes the condition of fixed displacement, stemming from the corrections at order $\mathcal{O}\bigl(\ell_0^{-2}\bigr)$ in Eqs.~(\ref{eq:fgdef}). Hence, already at this order that does not even resolve the effect of non-zero differential tension $\delta$, the governing equation differs from Euler's Elastica equation.}. We therefore truncate the expansion at fourth order to obtain a description of the epithelium in the spirit of a Landau theory. Not only do nonlinear elastic terms arise at this order of truncation, but nonlocal terms appear, too: the theory is \emph{not} elastic~\cite{libai}, since the energy depends not only on strain (i.e. curvature), but also on its (spatial) derivatives, introducing a nonlocal dependence on strain.

To obtain the governing equation, we vary the truncated expansion (\ref{eq:Lagrangian}) with respect to $\psi$, noting that $\Lambda$ is a constant since the trapezoidal cells are required to match up exactly~\footnote{In Ref.~\cite{krajnc15}, the energy was expressed in terms of $\psi$ and $\ell$, which were then both varied, while they are in fact related by $\ell\sec{\phi(\psi)}=\text{const.}$ for matching cells. If the lateral sides are not required to match up exactly, corrections to the final term in Eq.~(\ref{eq:singlecellE}) would have to be introduced, however, to ensure a consistent description of the adhesion between neighboring cells. Notwithstanding this, minimising Eq.~(\ref{eq:nondime}) with respect to $\lambda$ leads to $e(\phi)=2\ell_0\sqrt{\sec{\phi}+\delta\tan{\phi}}$, with $e''(0)=\bigl(2-\delta^2\bigr)\ell_0/4$. For this reason, if the lateral sides of the cells need not match up, the flat epithelium is unstable to small perturbations if $\smash{\delta>\sqrt{2}}$, as obtained in Ref.~\cite{krajnc15}}. After a considerable amount of algebra, we find
\begin{align}
\ddddot{\psi}=6\Xi^2\ddot{\psi}-\dfrac{3\Delta\Xi\dot{\psi}\ddot{\psi}}{1+\mu\cos{\psi}}+\dfrac{15+27\mu\cos{\psi}}{4(1+\mu\cos{\psi})}\dot{\psi}^2\ddot{\psi}+\dfrac{\mu\Xi^4\sin{\psi}}{1+\mu\cos{\psi}}\left[24-\dfrac{3}{\Xi^2}\dot{\psi}^2-\dfrac{1}{\Xi^4}\left(\dfrac{23}{16}\dot{\psi}^4-\dfrac{3}{2}\ddot{\psi}^2-2\dot{\psi}\dddot{\psi}\right)\right],\label{eq:goveq}
\end{align}
wherein $\Delta=\delta\Lambda^2$, subject to the boundary conditions
\begin{subequations}\label{eq:bc}
\begin{align}
&\psi(0)=\psi(1)=0,&& \ddot{\psi}(0)=\ddot{\psi}(1)=0,
\end{align}
and the integral condition
\begin{align}
\int_0^1{\cos{\psi}\left(1+\dfrac{\dot{\psi}^2}{8\Xi^2}+\dfrac{41\dot{\psi}^4}{1920\Xi^4}+\dfrac{\ddot{\psi}^2}{40\Xi^4}+\dfrac{\dot{\psi}\dddot{\psi}}{240\Xi^4}\right)\mathrm{d}\sigma}=\Lambda(1-D). \label{eq:displ}
\end{align}
\end{subequations}
\end{widetext}
The last condition imposes the fixed end-to-end shortening of the epithelium. These equations have a trivial solution $\psi=0$, $\Lambda=(1-D)^{-1}$, corresponding to the compressed but unbuckled state of the epithelium.

We note that, although Eq.~(\ref{eq:goveq}) only depends on $\delta$ and~$\Lambda$ through their agglomerate $\Delta$, a separate dependence on~$\Lambda$ arises in condition (\ref{eq:displ}). Minimising the energy of buckled solutions of Eqs. (\ref{eq:goveq},\ref{eq:bc}) with respect to $\Lambda$ finally determines $\Lambda$.
\section{Buckling Analysis}
In this section, we analyze the buckling behavior of the epithelium, first determining the threshold for buckling analytically and then discussing the post-buckling behavior using a weakly nonlinear analysis of the governing equations.

The buckling analysis naturally divides into two parts: we first seek buckled configurations of small amplitude for each value of $\Lambda$, and then minimize the energy of these configurations with respect to $\Lambda$. 

\subsection{Solution of the buckling problem}
The form of the trivial solution and of condition (\ref{eq:displ}) suggest that the appropriate small parameter for the first part of the analysis is
\begin{align}
\varepsilon^2=1-\Lambda(1-D). \label{eq:defeps}
\end{align}
We therefore expand
\begin{align}
\psi(\sigma)=\varepsilon\Bigl(\psi_0(\sigma)+\varepsilon\psi_1(\sigma)+\varepsilon^2\psi_2(\sigma)+\mathcal{O}\bigl(\varepsilon^3\bigr)\Bigr), 
\end{align}
and write $\mu=\mu_0+\varepsilon\mu_1+\varepsilon^2\mu_2+\mathcal{O}\bigl(\varepsilon^3\bigr)$. It is important to note that, while the governing equations derived above are only valid in the limit $\Xi\gg 1$, this parameter is not an asymptotic parameter for the buckling analysis. It will however be useful to introduce
\begin{align}
\xi=\dfrac{\pi}{\Xi}\ll 1. 
\end{align}
Next, we solve Eq.~(\ref{eq:goveq}), subject to the boundary and integral conditions (\ref{eq:bc}), order-by-order. 

\subsubsection{Solution at order $\mathcal{O}(\varepsilon)$}
At lowest order, the problem becomes
\begin{align}
\ddddot{\psi}_0-6\Xi^2\ddot{\psi}_0-\dfrac{24\Xi^4\mu_0}{1+\mu_0}\psi_0=0, 
\end{align}
subject to $\psi_0(0)=\psi_0(1)=0$, $\ddot{\psi}_0(0)=\ddot{\psi}_0(1)=0$. The lowest eigenvalue of this problem is
\begin{align}
\mu_0=\dfrac{z}{1-z},\quad\text{where }z=\dfrac{\xi^2}{4}\left(1+\dfrac{\xi^2}{6}\right),
\end{align}
the corresponding solution for $\psi_0$ being
\begin{align}
\psi_0(\sigma)=\Psi_0\sin{\pi\sigma}. 
\end{align}

\subsubsection{Solution at order $\mathcal{O}\bigl(\varepsilon^2\bigr)$}
At next order, upon substituting for $\mu_0$,
\begin{align}
&\dddot{\psi}_1-6\Xi^2\ddot{\psi}_1-24z\Xi^4\psi_1\nonumber\\
&\hspace{10mm}=24\mu_1\Xi^4(1-z)^2\psi_0-3\Delta\Xi(1-z)\dot{\psi}_0\ddot{\psi}_0.
\end{align}
\begin{widetext}
\noindent subject to $\psi_1(0)=\psi_1(1)=0$, $\ddot{\psi}_1(0)=\ddot{\psi}_1(1)=0$. These conditions imply that $\mu_1=0$, which is the usual result for the supercritical pitchfork bifurcation expected for this buckling problem. Thence
\begin{align}
\psi_1(\sigma) = \Psi_1\sin{\pi\sigma} + \Delta \Psi_0^2\left(\dfrac{\xi}{12}-\dfrac{13\xi^3}{144}\right)\sin{2\pi\sigma}+\mathcal{O}\bigl(\xi^5\bigr).
\end{align}

\subsubsection{Solution at order $\mathcal{O}\bigl(\varepsilon^3\bigr)$}
Finally, upon substituting for $\mu_0$ and $\mu_1=0$,
\begin{align}
\ddddot{\psi}_2- 6\Xi^2\ddot{\psi}_2-24z\Xi^4\psi_2&=24\Xi^4\mu_2(1-z)^2\psi_0-3\Delta\Xi(1-z)\Bigl(\dot{\psi}_0\ddot{\psi}_1+\ddot{\psi}_0\dot{\psi}_1\Bigr)+\left(\tfrac{15}{4}+3z\right)\dot{\psi}_0^2\ddot{\psi}_0\nonumber\\ 
&\hspace{40mm}-4z(1-3z)\Xi^4\psi_0^3-z\,\psi_0\left(3\Xi^2\dot{\psi}_0^2-\tfrac{3}{2}\ddot{\psi}_0^2-2\dot{\psi}_0\dddot{\psi}_0\right), 
\end{align}
subject to $\psi_2(0)=\psi_2(1)=0$, $\ddot{\psi}_2(0)=\ddot{\psi}_2(1)=0$. After a considerable amount of algebra, we obtain
\begin{align}
\mu_2=\Psi_0^2\left[\dfrac{\xi^2}{32}+\left(\dfrac{17}{384}-\dfrac{\Delta^2}{96}\right)\xi^4\right]+\mathcal{O}\bigl(\xi^5\bigr).
\end{align}
and thence
\begin{align}
\psi_2(\sigma)=\Psi_2\sin{\pi\sigma}+\Psi_0\Psi_1 \Delta\left(\dfrac{\xi}{6}-\dfrac{13\xi^3}{72}\right)\sin{2\pi\sigma}+\Psi_0^3\left(\dfrac{1}{192}+\dfrac{4\Delta^2-9}{256}\xi^2-\dfrac{36\Delta^2-37}{768}\xi^4\right)\sin{3\pi\sigma}+\mathcal{O}\bigl(\xi^5\bigr).
\end{align}
\end{widetext}
\subsubsection{Calculation of the amplitudes}
The constants $\Psi_0,\Psi_1,\Psi_2$ left undetermined by the above calculation are obtained by expanding both sides of the integral condition (\ref{eq:displ}). Solving order-by-order, we obtain
\begin{align}
\Psi_0&=\dfrac{2}{\sqrt{1-z}}, &\Psi_1=0,
\end{align}
where we have chosen $\Psi_0>0$ without loss of generality. The result $\Psi_1=0$ is to be expected for a supercritical bifurcation; in fact, in a standard elastic buckling problem, one would have $\psi_1\equiv 0$; here, a non-zero $\psi_1$ is required because of the symmetry breaking resulting from the term proportional to $\dot{\psi}^3$ in the Lagrangian~(\ref{eq:Lagrangian}). Further, we obtain \enlargethispage{4mm}
\begin{align}
\Psi_2=\dfrac{1}{4}+\left(\dfrac{5}{32}-\dfrac{\Delta^2}{36}\right)\xi^2+\left(\dfrac{149}{7680}+\dfrac{37\Delta^2}{864}\right)\xi^4+\mathcal{O}\bigl(\xi^5\bigr). 
\end{align}

\subsection{Minimization of the Energy}
In the second part of the buckling analysis, we determine the buckling threshold, and then analyze the post-buckling behavior. Substituting for $\Lambda$ using Eq.~(\ref{eq:defeps}) in the energy term in Eq.~(\ref{eq:Lagrangian}) and expanding, the energy of the buckled configuration is
\begin{align}
\left(\dfrac{1}{1-D}+1-D\right)+\left(\dfrac{1-D}{1-z}-\dfrac{1}{1-D}\right)\varepsilon^2+\mathcal{O}\bigl(\varepsilon^3\bigr), 
\end{align}
wherein the first bracketed term is the energy of the trivial solution $\psi=0$, $\Lambda=(1-D)^{-1}$. Accordingly, buckled configurations become energetically favorable if
\begin{align}
\dfrac{1-D}{1-z}-\dfrac{1}{1-D}<0\quad\Longleftrightarrow\quad D>D_\ast\equiv 1-\sqrt{1-z}. 
\end{align}
In particular, the buckling threshold is independent of the differential tension $\delta$.

We are left to determine the value of $\Lambda$ that minimizes the energy of the buckled configuration. This is equivalent with relating $\varepsilon$ to the excess compression $d=D-D_\ast>0$. We therefore write $\varepsilon=\varepsilon_0d^{1/2}+\mathcal{O}(d)$, and obtain an expansion of the energy in $d\ll 1$,
\begin{align}
\dfrac{\mathcal{E}}{\ell_0^2}=\left(C\varepsilon_0^4-\dfrac{2\varepsilon_0^2}{1-z}\right)d^2+\mathcal{O}\bigl(d^{5/2}\bigr), 
\end{align}
where
\begin{align}
C=1+\dfrac{3\xi^2}{16}+\left(\dfrac{11}{64}-\dfrac{\delta^2}{48}\right)\xi^4+\mathcal{O}\bigl(\xi^5\bigr). 
\end{align}
The energy is thus minimized when $\varepsilon_0=\bigl(C(1-z)\bigr)^{-1/2}$. Substituting this result into the expression for $\mu_2$, we finally obtain
\begin{align}
\mu\sim\mu_0(\xi)+\left(\dfrac{\xi^2}{8}+\dfrac{83-16\delta^2}{384}\xi^4+\mathcal{O}\bigl(\xi^5\bigr)\right)(D-D_\ast). 
\end{align}
This is the main result of our asymptotic analysis of the buckling: the force required to compress the epithelium decreases with increasing differential tension $\delta>0$.

In general, the buckled configuration features both energetically favorable regions ($\dot{\psi}>0$ if $\delta>0$) and unfavorable regions ($\dot{\psi}<0$). Still, this result shows that buckling overall is facilitated if $\delta>0$ compared to the $\delta=0$ case.

\begin{figure*}[t]
\includegraphics{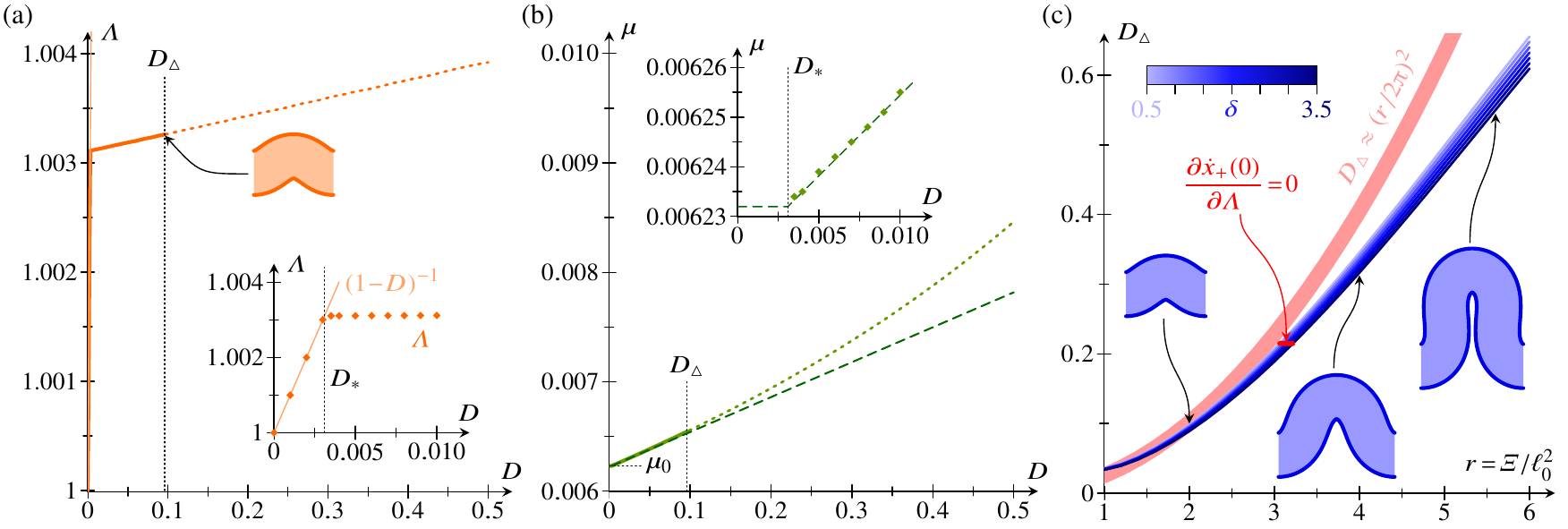}
\caption{Numerical buckling results. (a) Plot of $\Lambda$ against relative compression $D$. Above a critical compression $D_\triangle$ (buckled shape for $D=D_\triangle$ shown), solution shapes at the energy minimum begin to self-intersect (dotted line for $D>D_\triangle$). Inset: zoomed plot of $\Lambda$ (filled marks) against~$D$ close to the buckling threshold $D_\ast$. (b)~Plot of compressive force $\mu$ against relative compression $D$, showing numerical results (solid line and dotted line for $D>D_\triangle$) in agreement with asymptotic results (dashed line). Inset: zoomed plot of $\mu$ against $D$ close to $D_\ast$. Parameter values for numerical calculations: $\Xi=20$, $\delta=1$, $\ell_0=\sqrt{10}$. (c)~Critical compression $D_\triangle$ against $\smash{r=\Xi/\ell_0^2}$, for different values of~$\delta$, at fixed $\ell_0=\sqrt{10}$, and approximation (\ref{eq:DTapprox}) thereof. Insets show buckled shapes at $D=D_\triangle$ and $\delta=1$, for different values of $\Xi$.}
\label{fig:results} 
\end{figure*}

\section{Post-buckling behavior}
While asymptotic analysis can describe the deformations of the epithelium just beyond the buckling threshold, larger compressions must be studied numerically. We solve the governing equation (\ref{eq:goveq}), complemented by the boundary and integral conditions (\ref{eq:bc}), numerically using the boundary-value-problem solvers \texttt{bvp4c}, \texttt{bvp5c} of \textsc{Matlab} (The MathWorks, Inc.) and the continuation package \textsc{auto}~\cite{auto}. 

\subsection{Transition to Constricted Cells}
For the numerical solution, we fix $\Xi$ and $\delta$, and obtain solutions for different values of $\Lambda$. By interpolation, we determine the value of $\Lambda$ that minimizes the energy (Fig.~\ref{fig:results}a). Thence we obtain the corresponding value of the compressive force $\mu$ (Fig.~\ref{fig:results}b), in agreement with the asymptotic results of the previous section. This also validates our numerical implementation of the system.

There is, however, one extra constraint that has not been incorporated into the continuum equations: the constraint, related to the transition to constricted cells that we have briefly discussed before, that the lateral sides of the trapezoidal cells cannot self-intersect. At the level of the continuum description, this constraint translates to the condition that the apical and basal surfaces of the epithelium cannot self-intersect. The apical and basal surfaces of the cell sheet are described by the curves
\begin{align}
x_\pm&=x\mp\ell\sin{\psi},&y_\pm&= y\pm\ell\cos{\psi},\label{eq:xypm}
\end{align}
where $\ell = \ell_0\Lambda\cos{\phi}$ is the local thickness of the cell sheet. It is important to note that Eqs.~(\ref{eq:xypm}) are exact equations since the Kirchhoff `hypothesis' of the analysis of slender elastic  structures, the asymptotic result~\cite{audolypomeau} that the normal to the undeformed midline remains normal in the deformed configuration, is an exact result in the discrete model that underlies our continuum theory. (For this same reason, an analogous analysis for an elastic object beyond asymptotically small deformations would, rather more intricately, require solving for the stretches in each parallel to the midline.)

Numerically, we find that, as $D$ is increased at fixed $\Lambda$, the shapes of minimal energy self-intersect above a critical compression $D_\triangle$. The numerical solutions also reveal that self-intersections first arise at $\sigma=0$, when $\dot{x}_+=0$ there. Expanding this condition using $\psi(0)=0$ and $\dot{\phi}(0)=0$, of which the latter follows from Eq.~(\ref{eq:phiexp}) by symmetry, 
\begin{align}
\dfrac{f(0)}{\Lambda}-\Lambda\dfrac{\ell_0^2}{\Xi}\dot{\psi}(0)\cos{\phi(0)}=0, \label{eq:diff}
\end{align}
where $f$ is defined as in Eq.~(\ref{eq:fdef}). We note that, with this condition, an explicit dependence on both $\ell_0$ and $\Xi$ has arisen for the first time in our analysis.

\begin{figure*}[t]
\includegraphics{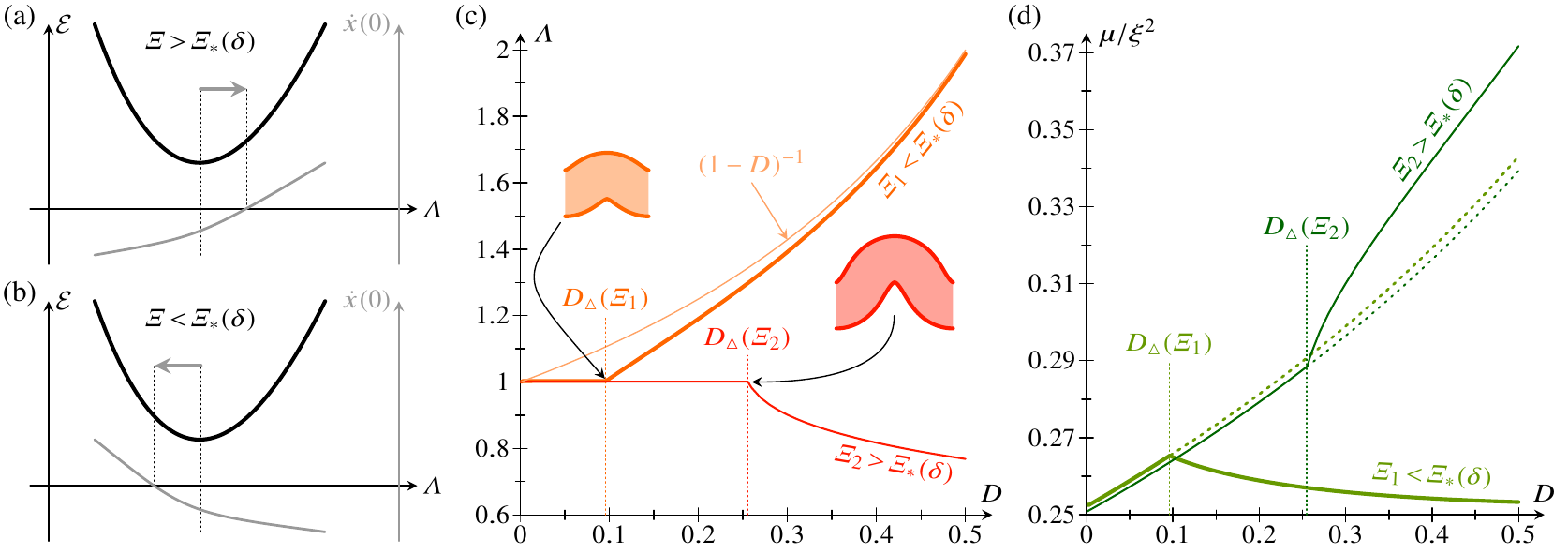}
\caption{Buckling for $D>D_\triangle$. Two scenarios are possible: (a)~If $\partial\dot{x}(0)/\partial\Lambda>0$ or $\Xi<\Xi_\ast(\delta)$, buckled shapes with increased $\Lambda$ do not self-intersect. (b)~If $\partial\dot{x}(0)/\partial\Lambda<0$ or $\Xi>\Xi_\ast(\delta)$, buckled shapes with decreased $\Lambda$ do not self-intersect. Numerical results: (c)~Plot of $\Lambda$ (for buckled shapes of minimal energy without self-intersections) against relative compression $D$, for $\Xi_1<\Xi_\ast$ (thick lines) and $\Xi_2>\Xi_\ast$ (thin lines). For $\Xi<\Xi_\ast$, $\Lambda\rightarrow(1-D)^{-1}$ for $D>D_\triangle$. Insets show buckled shapes at $D=D_\triangle$. (d)~Corresponding plots of scaled compressive force~$\mu/\xi^2$ against $D$. Dotted lines for $D>D_\triangle$ correspond to self-intersecting shapes at the energy minimum. Parameter values for numerical calculations: $\delta=1$, $\Xi_1=20$, $\Xi_2=34$, $\ell_0=\sqrt{10}$.}
\label{fig:results2} 
\end{figure*}

\subsubsection*{Estimating the critical compression $D_\triangle$}
The numerical data in Fig.~\ref{fig:results}a,b suggest that the asymptotic results of the previous section can approximate the buckling behavior of the epithelium well up to compressions as large as $D_\triangle$. We therefore use our asymptotic results to estimate the critical compression $D_\triangle$. For this purpose, we treat $r=\smash{\Xi/\ell_0^2}$ as an $\mathcal{O}(1)$ quantity. Then, using $\Lambda=(1-z)^{-1/2}+\mathcal{O}(d)$,
\begin{align}
\dfrac{f(0)}{\Lambda}-\Lambda\dfrac{\ell_0^2}{\Xi}\dot{\psi}(0)\cos{\phi(0)}=\sqrt{1-z}-\dfrac{\pi\varepsilon_0\Psi_0}{r\sqrt{1-z}}d^{1/2}+\mathcal{O}(d),
\end{align}
whence, to lowest order in $\xi$,
\begin{align}
D_\triangle\approx\dfrac{r^2}{4\pi^2}. \label{eq:DTapprox}
\end{align}
This approximation is not itself an asymptotic result, yet, for small enough values of $r$, it compares well to numerical estimates of $D_\triangle$ obtained by a bisection search (Fig.~\ref{fig:results}c). The numerical results also show that, at fixed $\Xi$, $D_\triangle$ decreases with increasing $\delta$, with relative variations of about $10\%$ for the range of $\delta$ under consideration.

We also find numerically that, at large values of $r$, steric interactions between different parts of the cell sheet become important before $D$ reaches $D_\triangle$. A detailed analysis of these interactions is beyond the scope of this discussion.

We have tacitly assumed that, at fixed $\Xi$ and at fixed $D>D_\ast$, the energy $\mathcal{E}$ has a single local minimum as a function of $\Lambda$. This is indeed the case for small enough values of $\delta$, but fails at compressions $D>D_\triangle$ as~$\delta$ is increased, so this possibility is not of direct relevance to the present discussion. Interestingly, eigenmodes (buckled solutions with zero force) of the epithelium arise at large $\delta$. These eigenmodes are not energy minimizers, but, for completeness, we discuss these solutions in Appendix~\ref{appB}. 

\subsection{Buckled shapes for $\vec{D>D_\triangle}$}
As $D$ is increased beyond $D_\triangle$, we might expect fans of constricted cells to expand around the trough and (later) the crest of the buckled shape, but deriving the equations describing these fans and solving for these shapes is beyond the scope of the present discussion. 

Here, we note simply that, for $D>D_\triangle$, buckled shapes without self-intersections can be found. Two scenarios can be envisaged \emph{a priori}, depending on the sign of $\partial\dot{x}(0)/\partial\Lambda$ at the energy minimum (Fig.~\ref{fig:results2}a,b): if $\partial\dot{x}(0)/\partial\Lambda>0$, buckled solutions without self-intersection arise as $\Lambda$ is increased; if $\partial\dot{x}(0)/\partial\Lambda>0$, such shapes are found as $\Lambda$ is decreased. Interestingly, both of these possibilities arise in the system: there exists a critical value $\Xi=\Xi_\ast(\delta)$ at which $\partial\dot{x}(0)/\partial\Lambda=0$. The first possibility occurs in the case $\Xi<\Xi_\ast$, and the second one in the case $\Xi>\Xi_\ast$ (Fig.~\ref{fig:results2}c).

If $\Xi<\Xi_\ast$, the buckling amplitude decreases for these solutions as $D$ is increased beyond $D_\triangle$: $\Lambda$ tends to its value $(1-D)^{-1}$ in the unbuckled, compressed configuration, while $\mu$ decreases (Fig.~\ref{fig:results2}c,d). By contrast, if $\Xi>\Xi_\ast$, the buckling amplitude is increased: as $D$ increases and $\Lambda$ decreases, $\mu$ increases faster than in the self-intersecting configurations at the energy minimum (Fig.~\ref{fig:results2}c,d). 

While these qualitative considerations cannot capture the exact mechanics of the fans of constricted cells near the through and crest of the buckled shape, we expect them to give a qualitative indication of the buckling behavior as $D$ is increased just beyond $D_\triangle$. For larger values of $D$, there are more intricate possibilities: there are in general two values of~$\Lambda$ such that $\dot{x}(0)=0$, on either side of the energy minimum. One of these solutions defines the branch shown in Fig.~\ref{fig:results2}c, but the second solution may become energetically favorable over the first one as $D$ is increased sufficiently. This actually happens on the branch with $\Xi=\Xi_1<\Xi_\ast$ in Fig.~\ref{fig:results2}c at $D\approx 0.44$, but, for the second solution, different parts of the cell sheet start to touch before this value of $D$ is reached and hence we do not pursue this further here.

\section{Conclusion}
In this paper, we have derived, by taking a rigorous asymptotic limit, the continuum limit of a simple discrete differential-tension model of a two-dimensional epithelium. If the expansion is carried to high enough order for the differential tension between the apical and basal sides of the epithelium to arise in the energy, nonelastic terms that are nonlocal in the strains appear. This is the key lesson to be drawn from taking the continuum limit. We have gone on to use this continuum model to study the buckling of the epithelium under imposed confinement, showing how, post-buckling, the compressive force is reduced with increasing differential tension. A second buckling transition occurs when constricted cells start to form near the troughs and crests of the buckled shape; we have discussed the behavior close to this transition qualitatively. Taking the analysis of the buckling behavior of epithelium in this continuum framework beyond the transition to constricted cells is the key challenge for future work on this problem. 

Possible extensions of the continuum framework include mimicking the setup of studies of the discrete model~\cite{storgel16,hocevar12,rauzi15} by coupling the epithelium to an elastic substrate or incorporating fixed-volume constraints for a closed one-dimensional epithelium. The question of how to extend the continuum model to describe a two-dimensional epithelium also remains open. In particular, how are the deviations from elasticity affected by the increase of the dimensionality of the system?

Cell sheet deformations during development commonly feature large \emph{geometric} deformations, but the \emph{elastic} deformations can remain small provided that the deformed geometry remains close to the intrinsic geometry that is generally different from the initial geometry because of cell shape changes, cell intercalation, and related processes. For this reason, developmental events as intricate as the inversion process of the green alga \emph{Volvox} can be modelled quantitatively using a Hookean shell theory~\cite{hohn15,haas18}. By contrast, large deformations of many a biological material are not in general described well by neo-Hookean constitutive equations, although other families of hyperelastic constitutive equations predict behavior in quantitative agreement with experimental data for brain and fat tissues~\cite{mihai15,mihai17,budday17}. However, and in spite of the ubiquity of these elastic models, in particular in the modelling of the folds of the cerebrum~\cite{richman75,tallinen13,manyuhina14,budday15,tallinen16}, it was pointed out very recently that the folding of the cerebellum is fundamentally inconsistent with the differential-growth hypothesis~\cite{engstrom18}: in the cerebellum, the oscillations of the thicknesses of the core and the growing cortex are out-of-phase, while elastic bilayer instabilities lead to in-phase oscillations~\cite{engstrom18}. All of this emphasizes the need for a deeper understanding of how continuum models relate to properties of structures at the cell level. By explicitly showing how both nonlinear and nonlocal elastic terms arise in the continuum limit of a simple discrete model and impact on its behavior, the present analysis has taken a first step in this direction.

\vspace{-3mm}

\begin{acknowledgments}
We thank Matej Krajnc for discussions of Ref.~\cite{krajnc15} at an early stage of this work, and are grateful for support from the Engineering and Physical Sciences Research Council (Established Career Fellowship EP/M017982/1, REG; Doctoral Prize Fellowship, PAH), Wellcome Trust Investigator Award 207510/Z/17/Z, the Schlumberger Chair Fund, and Magdalene College, Cambridge (Nevile Research Fellowship, PAH).
\end{acknowledgments}

\vspace{6mm}
\appendix \section{Derivation of Eq.~(\ref{eq:dpsi})} 
\renewcommand{\theequation}{A\arabic{equation}}\label{appA}
\setcounter{equation}{0}
In this appendix, we relate $\phi$ and $\psi$ by deriving Eq.~(\ref{eq:dpsi}) used in the main text. At the same time, we verify that this expansion is indeed consistent at all orders. Taylor expanding the right-hand side of Eq. (\ref{eq:psigeo}),
\begin{widetext}
\begin{align}
\psi(s+ks_0)-\psi(s)&=\phi(s)+2\left(\sum_{j=1}^{k-1}{\phi(s+js_0)}\right)+\phi(s+ks_0)=2k\phi(s)+\sum_{n=1}^\infty{\dfrac{s_0^n}{n!}\phi^{(n)}(s)\left\{k^n+2\sum_{j=1}^{k-1}{j^n}\right\}}\nonumber\\
&\hspace{25mm}=2k\phi(s)+\sum_{n=1}^\infty{\dfrac{s_0^n}{n!}\phi^{(n)}(s)\left\{k^n+\dfrac{2}{n+1}\sum_{j=0}^n{(-1)^j\binom{n+1}{j}\mathcal{B}_j(k-1)^{n+1-j}}\right\}},
\end{align}
where we have used Faulhaber's formula~\cite{conway} to expand the sum of powers of integers, and where $\mathcal{B}_0=1,\mathcal{B}_1=-\frac{1}{2},\dots$ denote the Bernoulli numbers (of the first kind)~\cite{bernoulli}. Expanding $\smash{(k-1)^{n+1-j}}$ using the binomial theorem and simplifying the binomial coefficients,
\begin{align}
\psi(s+ks_0)-\psi(s)= 2k\phi(s)+\sum_{n=1}^\infty{\dfrac{\phi^{(n)}(s)}{n!\ell_0^n}A(k,n)},\label{eq:dpsiA}
\end{align}
where we have introduced
\begin{align}
A(k,n)&=k^n+2(-1)^{n+1}n!\sum_{j=0}^n{\sum_{i=0}^{n+1-j}{\dfrac{(-1)^i\mathcal{B}_j k^i}{i!j!(n+1-i-j)!}}}=\sum_{i=0}^{n+1}{a_i(n)k^i},
\end{align}
wherein the coefficients $a_0,a_1,\dots,a_{n+1}$ depend on $n$. We notice in particular that
\begin{align}
a_0&=0,&&a_n=0,&&a_{n+1}=\dfrac{2}{n+1},
\end{align}
of which the last two are obtained by direct computation, and the first one follows using an identity of Bernoulli numbers~\cite{bernoulli},
\begin{align}
\sum_{j=0}^{n}{\binom{n+1}{j}\mathcal{B}_j}=0\qquad\mbox{for }n=1,2,\dots. \label{eq:Bid}
\end{align}
Moreover, for $i=1,2,\dots,n-1$,
\begin{align}
a_i(n)=2(-1)^{n+1-i}\dfrac{n!}{i!}\sum_{j=0}^{n+1-i}{\dfrac{\mathcal{B}_j}{j!(n+1-i-j)!}}. 
\end{align}
Accordingly,
\begin{align}
A(k,n)=\dfrac{2k^{n+1}}{n+1}+2(-1)^{n+1}n!\sum_{i=1}^{n-1}{(-1)^i\dfrac{k^i}{i!}\left\{\sum_{j=0}^{n+1-i}{\dfrac{\mathcal{B}_j}{j!(n+1-i-j)!}}\right\}}. 
\end{align}
Upon inverting the order of summation, Eq.~(\ref{eq:dpsiA}) becomes
\begin{align}
&\psi(s+ks_0)-\psi(s)=2k\phi(s)+\sum_{n=1}^\infty{\dfrac{\phi^{(n)}(s)}{n!\ell_0^n}\left(\dfrac{2k^{n+1}}{n+1}\right)}+\sum_{i=1}^\infty{\sum_{n=i+1}^\infty{2(-1)^{n+1-i}\dfrac{\phi^{(n)}(s)}{\ell_0^n}}\dfrac{k^i}{i!}\left\{\sum_{j=0}^{n+1-i}{\dfrac{\mathcal{B}_j}{j!(n+1-i-j)!}}\right\}}.
\end{align}
Finally, upon relabelling indices in the first summation and changing variables $n\longmapsto m=n+1-i$ in the last summation,
\begin{align}
&\psi(s+ks_0)-\psi(s)=\sum_{i=1}^\infty{\dfrac{k^i}{i!\ell_0^i}\left\{2\ell_0\phi^{(i-1)}(s)+\sum_{m=2}^\infty{\dfrac{2(-1)^m}{\ell_0^{m-1}}}\phi^{(i-1+m)}(s)\left(\sum_{j=0}^m{\dfrac{\mathcal{B}_j}{j!(m-j)!}}\right)\right\}}. 
\end{align}
\end{widetext}
But, rearranging Eq.~(\ref{eq:Bid}),
\begin{align}
\sum_{j=0}^m{\dfrac{\mathcal{B}_j}{j!(m-j)!}}=\dfrac{\mathcal{B}_m}{m!}\qquad\mbox{for }m=2,3,\dots.
\end{align}
Since $\mathcal{B}_n=0$ for odd $n>1$, and using $\mathcal{B}_0=1$, we finally obtain
\begin{align}
\psi(s+ks_0)-\psi(s)=\sum_{i=1}^\infty{\dfrac{k^i}{i!\ell_0^i}\left\{\sum_{m=0}^\infty{\dfrac{2\mathcal{B}_{2m}}{(2m)!}\dfrac{\phi^{(i-1+2m)}(s)}{\ell_0^{2m-1}}}\right\}}.  
\end{align}
Comparing this to the Taylor expansion of the left-hand side,
\begin{align}
\psi(s+ks_0)-\psi(s)=\sum_{i=1}^\infty{\dfrac{k^i}{i!\ell_0^i}\psi^{(i)}(s)},  
\end{align}
we deduce that the expansion is consistent at all orders, with, in particular,
\begin{align}
\psi'(s)=\sum_{m=0}^\infty{\psi_m\dfrac{\phi^{(2m)}(s)}{\ell_0^{2m-1}}}\qquad\text{where }\psi_m=\dfrac{2\mathcal{B}_{2m}}{(2m)!},
\end{align}
which is Eq.~(\ref{eq:dpsi}). As noted in the main text, we are not aware of a closed form for the inverted series that expresses $\phi$ as a function of the derivatives of $\psi$. Formally, inverting Eq.~(\ref{eq:dpsi}) gives
\begin{align}
\phi(s)=\sum_{m=0}^\infty{\phi_m\dfrac{\psi^{(2m+1)}(s)}{\ell_0^{2m+1}}},
\end{align}
where the coefficients $\phi_0,\phi_1,\dots$ are determined recursively by $\phi_0\psi_0=1$ and
\begin{align}
\sum_{j=0}^m{\phi_j\psi_{m-j}}=0\qquad\text{for }m=1,2,\dots.
\end{align}
In agreement with Eq.~(\ref{eq:phiexp}), we find
\begin{align}
&\phi_0=\tfrac{1}{2}, &&\phi_1=-\tfrac{1}{24},&&\phi_2=\tfrac{1}{240},&&\dots.
\end{align}

\begin{figure}[b]
\includegraphics{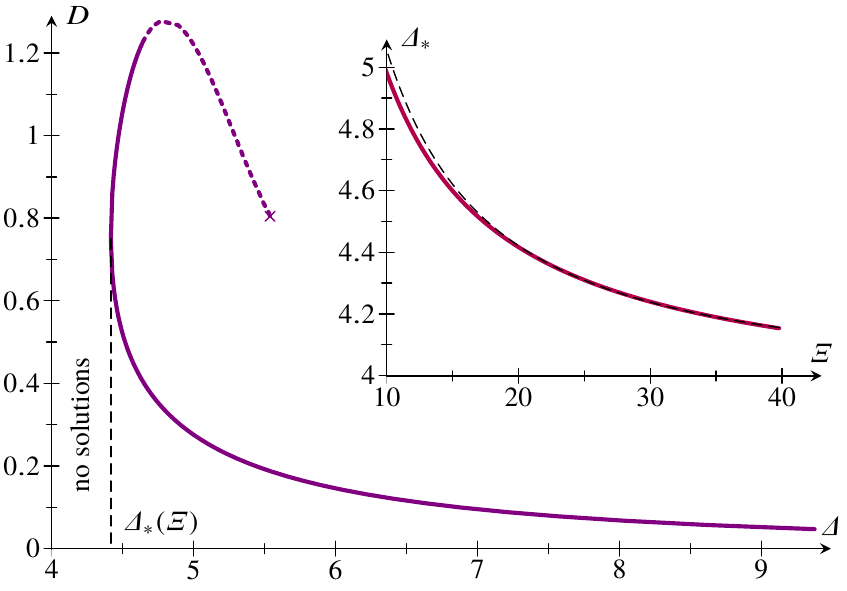}
\caption{Eigenmodes of a buckled epithelium: plot of relative end-to-end shortening $D$ against $\Delta$. Parameter value: $\Xi=20$. No eigenmodes were found for $\Delta<\Delta_\ast(\Xi)$. On the dashed part of the branch, $\mathcal{E}<2$. Continuation failed at the point marked $\times$. Inset: plot of $\Delta_\ast$ against $\Xi$ (solid line) and power-law fit (dashed line).} 
\label{fig:emodes}
\end{figure}

\section{Eigenmodes of the buckled epithelium} 
\renewcommand{\theequation}{B\arabic{equation}}\label{appB}
\setcounter{equation}{0}
\renewcommand{\thefigure}{B\arabic{figure}}
\setcounter{figure}{0}

Eigenmodes of the epithelium are non-zero solutions of the governing equation~(\ref{eq:goveq}) with $\mu=0$. They thus obey 
\begin{align}
\ddddot{\psi}=6\Xi^2\ddot{\psi}-3\Delta\Xi\dot{\psi}\ddot{\psi}+\dfrac{15}{4}\dot{\psi}^2\ddot{\psi},
\end{align}
subject to
\begin{align}
&\psi(0)=\psi(1)=0,&&\ddot{\psi}(0)=\ddot{\psi}(1)=0. 
\end{align}
To find eigenmodes numerically, we remove the trivial, zero solution by imposing a non-zero compression $D$ and varying this compression until a solution with $\mu=0$ is found.

Numerically, we obtain eigenmodes if $\Delta\geq\Delta_\ast$, but find no solutions if $\Delta<\Delta_\ast$, for some value $\Delta_\ast$ depending on $\Xi$ (Fig.~\ref{fig:emodes}). Some of these solutions have energy $\mathcal{E}<2$ (Fig.~\ref{fig:emodes}), lower than the energy of the uncompressed, flat solution; these are spontaneous buckled modes that arise in the absence of external forces, but, as is apparent from the corresponding values $D>1$ (Fig.~\ref{fig:emodes}), these solutions are unphysical. Plotting $\Delta_\ast$ against $\Xi$ (Fig.~\ref{fig:emodes}, inset) suggests that $\Delta_\ast$ approaches a constant value as $\Xi$ grows large. We observe that the numerical data are well approximated by a power-law $\Delta_\ast=c_1+c_2\Xi^{-5/4}$, where $c_1\approx 3.96$, $c_2\approx 19.5$ (Fig.~\ref{fig:emodes}, inset).

The `large' values of $\Delta$ and hence $\delta$ for these eigenmodes beckon a comment on the formal range of validity of the continuum model: stability of the underlying discrete model requires $\alpha,\beta\geq 0$ \cite{krajnc13}, and hence $\delta\leq\ell_0^2$. While the asymptotic expansion leading to the geometric relation (\ref{eq:phiexp}) was an expansion in the large parameter $\ell_0$, it did not involve $\delta$. By contrast, the expansion of the Lagrangian (\ref{eq:Lagrangian}), which did involve $\delta$, was an expansion in a different large parameter, $\Xi$. Hence large values of $\delta\lesssim\ell_0^2$ are indeed in the formal range of validity of the continuum model provided that~$\Xi$ is large enough.

\bibliography{epi}
\end{document}